\documentclass[aps,prd,onecolumn,showpacs,showkeys,amsmath,amssymb]{revtex4}
\usepackage{graphicx}
\usepackage{dcolumn}
\usepackage{bm}
\begin{document}
\title{Strong decays of $D_{sJ}(2317)$ and $D_{sJ}(2460)$}
\author{Wei Wei, Peng-Zhi Huang}
\affiliation{Department of Physics, Peking University, Beijing
100871, China}
\author{Shi-Lin Zhu}
\email{zhusl@th.phy.pku.edu.cn} \affiliation{Department of
Physics, Peking University, Beijing 100871, China}

\date{\today}

\begin{abstract}

With the identification of ($D_{sJ}(2317), D_{sJ}(2460)$) as the
($0^+$, $1^+$) doublet in the heavy quark effective field theory,
we derive the light cone QCD sum rule for the coupling of eta
meson with $D_{sJ}(2317) D_s $ and $D_{sJ}(2460) D_s^{*} $.
Through $\eta-\pi^0$ mixing we calculate their pionic decay
widths, which are consistent with the experimental values (or
upper limits). Combining the radiative decay widths derived by
Colangelo, Fazio and Ozpineci in the same framework, we conclude
that the decay patterns of $D_{sJ}(2317, 2460)$ strongly support
their interpretation as ordinary $c \bar s$ mesons.

\end{abstract}



\maketitle

\pagenumbering{arabic}

\section{Introduction}\label{sec1}

Since the discovery of $D_{sJ}(2317)$ \cite{babar1} and
$D_{sJ}(2460)$ \cite{cleo}, there have been lots of experimental
investigations of these two narrow resonances
\cite{belle1,belle2,belle3,belle4,focus,babar2,babar3,babar4,babar5}.
They have the natural spin-parity assignment as the $0^+,1^+$
charm-strange mesons from the observed final states.  Their masses
are about one hundred MeV lower than the quark model prediction
\cite{quark model}, which are really unexpected. Many theoretical
papers have been dedicated to the understanding of their
underlying structure
\cite{Bardeen,Nowak,lutz,molecule,tetra1,tetra2,tetra3,tetra4,tetra5,
tetra6,tetra7,tetra8,atom,deandrea,cahn,slz,lucha,hofmann,sadzi,
beci,lee,alz,lattice,colangelo4}. Proposed schemes include the
($0^+,1^+$) chiral partners of the $D_s, D_s^{*}$ doublet in heavy
quark effective field theory \cite{Bardeen,Nowak,lutz}, $DK$
molecules \cite{molecule}, four quark states
\cite{tetra1,tetra2,tetra3,tetra4,tetra5,tetra6,tetra7,tetra8},
$D\pi$ atom \cite{atom} and conventional $c\bar s$ states
\cite{Bardeen,Nowak,deandrea,cahn,slz,lucha,hofmann,sadzi,beci,lee}.

These two states are lower than the $DK$ and $D^{*}K$ thresholds
respectively. Their strong decays are isospin violating and occur
through two steps: $D_{sJ}(2317)\rightarrow D_s+\eta \rightarrow
D_s+\pi^0$ and $D_{sJ}(2460)\rightarrow D^{*}_s+\eta\rightarrow
D^{*}_s+\pi^0$. The second step is induced by the $\eta-\pi^0$
mixing due to the mass difference between $m_u$ and $m_d$
\cite{wise}. There have been some discussions of their strong and
radiative decays within the quark model
\cite{Bardeen,tetra1,godfrey,colangelo1,fayy,mehen,colangelo4}.
The strong decay widths from various approaches differ
significantly as can be seen from Table \ref{table1}. Their
radiative decay widths were calculated using light cone QCD sum
rule (LCQSR) not long ago \cite{colangelo2}. Very recently, the
branch ratios of the strong and radiative decays were measured
quite accurately by Belle Collaboration \cite{belle1,belle2} and
Babar Collaboration \cite{babar2,babar5}. In order to pin down the
underlying quark content of these narrow states, a reliable
calculation of their strong decay widths will be very helpful.

In this work, we assume ($D_{sJ}(2317), D_{sJ}(2460)$) as the
$c\bar s$ states and study their strong decays in the LCQSR
framework, which has been used extensively in extracting low-lying
hadron masses and coupling constants in the past decade (see Ref.
\cite{Colangelo3} for a review). This paper is organized as
follows. We calculate the coupling constant $g_{D_{s0}D_s\eta}$
and the strong decay width of $D_{sJ}(2317) \to D_s \pi^0$ through
$\eta-\pi^0$ mixing in Section \ref{sec2}. $D_{sJ}(2460)$ decay is
presented in Section \ref{sec3}. We compare our results with
experimental data and other theoretical approaches in literature
and summarize our results in Section \ref{sec4} . We collect the
light cone wave functions of the $\eta$ meson in the appendix.

\section{$D_{sJ}(2317)\rightarrow D_s+\eta \rightarrow D_s+\pi^0$}\label{sec2}

The amplitude of the strong decay $D_{sJ}(2317)\rightarrow
D_s+\eta$ can be defined as
\begin{equation}
\langle \eta(q)D_s(p)| D_{s0}(p+q)\rangle =
m_{D_{s0}}g_{D_{s0}D_s\eta}\; .
\end{equation}
We calculate the coupling constant $g_{D_{s0}D_s\eta}$ through
the following correlation function
\begin{equation}
F(p^2,(p+q)^2)=i \int d^4x \; e^{i p \cdot x} \langle \eta(q) |
T[J^\dagger_5(x) J_0(0)] |0\rangle \; ,
\end{equation}
where $J_0(x)=\bar c(x) s(x)$ and $J_5(x)=\bar c(x) i\gamma_5
s(x)$ are the interpolating currents of ${\bar D}_{s0}$ and ${\bar
D}_s$ respectively.

At the quark level, the correlation function can be expressed in
terms of the eta meson light cone wave functions after the
insertion of the charm quark propagator at the leading order
\begin{equation} \langle 0|T\{c(x)\bar{c}(0)\}|0\rangle =
i\hat{S}_c^0(x) = \int \frac{d^4k}{(2\pi)^4i}e^{-ikx}
\frac{\not\!k+m_c}{m_c^2-k^2-i\epsilon}\; .
\end{equation}
Now we have
\begin{equation}
F(p^2,(p+q)^2)= \int   \frac{d^4 k}{(2 \pi)^4}  \int d^4x \;
\frac{e^{i (p -k)\cdot x}}{m_c^2 -k^2} \big[m_c\langle \eta(q) |
\bar s(x) i\gamma_5 s(0) |0\rangle - i k^\alpha \langle \eta(q) |
\bar s(x) \gamma_{\alpha}s(0) |0\rangle \Big] \; .
\end{equation}

In order to include the contribution from the twist-four eta meson
light-cone wave functions, we need the three particle piece in the
charm quark propagator:
\begin{eqnarray}
\langle 0 |T\{c(x)\bar{c}(0)\}|0\rangle &=& i\hat{S}_c^0(x)
-ig_s\int\frac{d^4k}{(2\pi )^4}e^{-ikx} \int_0^1dv\left[ \frac
12\frac{\not\!k+m_c}{(m^2_c-k^2)^2} G^{\mu\nu}(vx)\sigma_{\mu\nu}
+\frac {1}{m_c^2-k^2}vx_\mu G^{\mu\nu}(vx)\gamma_{\nu} \right]~,
\label{32}
\end{eqnarray}
where $G_{\mu\nu}=G_{\mu\nu}^c\frac{\lambda^c}2$ with $\mbox{\rm
tr}(\lambda^a\lambda^b)=2\delta^{ab}$, and $g_s$ is the strong
coupling constant. The complete expression of $F(p^2,(p+q)^2)$ up
to twist-four reads:
\begin{eqnarray}\label{q1} \nonumber
F(p^2,(p+q)^2)&=&\int_0^1\frac{du}{m_c^2-(p+uq)^2}\Big \{
-F_{\eta}\varphi_{\eta}(u)p.q+m_cF_{\eta}\mu_{\eta}\varphi_{p}(u)\\\nonumber
&& -F_{\eta}m_{\eta}^2(1+\frac{m_c^2}{m_c^2-(p+uq)^2})G(u)+\frac14
F_{\eta}m_{\eta}^2\frac{A(u)p.q}{m_c^2-(p+uq)^2}[1+\frac{2m_c^2}{m_c^2-(p+uq)^2}]
\Big \}
\\
 && +F_{\eta}m_{\eta}^2 \int_0^1 dv\int {\cal{D}}\alpha_{i}
\frac{p.q}{m_c^2-(p+(\alpha_1+v\alpha_3)q)^2}\Big \{2v
(2\varphi_{\perp}-\varphi_{\parallel})-(2\varphi_{\perp}-\varphi_{\parallel}
+2\tilde{\varphi}_{\perp}-\tilde{\varphi}_{\parallel})\Big \}\; ,
\end{eqnarray}
where
\begin{equation}
G(u) =-\int_0^{u} du^{'} B(u^{'})\; .
\end{equation}
$\varphi_{\eta} (u),\varphi_{p}(u), B(u)$ etc are light cone
amplitudes of $\eta$ meson \cite{ballA,ballB}, which are collected
in the Appendix.  $F_\eta =-\frac{2}{\sqrt{6}}f_\eta $, with
$f_\eta $ defined as
\begin{equation}
\langle 0|\frac{1}{\sqrt{6}}\left( \bar u (0)\gamma_\mu \gamma_5
u(0) + \bar d (0)\gamma_\mu \gamma_5 d(0) -2\bar s (0)\gamma_\mu
\gamma_5 s(0) \right) | \eta (q) \rangle =i f_\eta q_\mu \; .
\end{equation}

At the phenomenological level, $F(p^2,(p+q)^2)$ can be expressed
as
\begin{equation}\label{q2}
F(p^2,(p+q)^2)=\frac{m_{D_{s0}}^2m_{D_s}f_{D_{s0}}f_{D_s}g_{D_{s0}D_s\eta}}
{(m_{D_s}^2-p^2)(m_{D_{s0}}^2-(p+q)^2)} +\cdots\;.
\end{equation}
The ellipse denotes the contribution from the continuum. The decay
constants $f_{D_{s0}}$ and $f_{D_s}$ are defined as
 \begin{eqnarray}
\langle 0 | J_5^{\dagger} | D_s\rangle&=& f_{D_s} m_{D_s} \; ,  \\
\langle D_{s0} | J_0 | 0\rangle&=& f_{D_{s0}} m_{D_{s0}}\; .
\end{eqnarray}

Applying the double Borel transformation with respect to $p^2$ and
$(p+q)^2$ to Eqs. (\ref{q1}) and (\ref{q2}) and invoking the
quark-hadron duality, we get the following sum rule:
\begin{eqnarray}\label{sr1} \nonumber
f_{D_{s0}}f_{D_s}g_{D_{s0}D_s\eta}&=&\frac{1}{m_{D_{s0}}^2m_{D_s}}
e^{\frac{m_{D_{s0}}^2+m_{D_s}^2}{2M^2}} \Bigg\{M^2
[e^{-\frac{m_c^2}{M^2}} - e^{-\frac{s_0}{M^2}}]
\Big[\frac{1}{2}M^2F_{\eta}\varphi^{'}_{\eta}(u_0)
+m_cF_{\eta}\mu_{\eta}\varphi_p({u_0})\\\nonumber
 &&-F_{\eta}m^2_{\eta}G(u_0)
-\frac 18 F_{\eta}m^2_{\eta}A^{'}(u_0)+\frac 12
F_{\eta}m^2_{\eta}(2I_1(2\varphi_{\perp}-\varphi_{\parallel})-
I_2(2\varphi_{\perp}-\varphi_{\parallel}
+2\tilde{\varphi}_{\perp}-\tilde{\varphi}_{\parallel}))\Big]\\
&& -e^{-\frac{m_c^2}{M^2}}F_{\eta}m^2_{\eta}m^2_c\Big[G(u_0)
+\frac18A^{'}(u_0)\Big]\Bigg\}_{u_0=1/2}\; .
\end{eqnarray}
The functions $I_1$ and $I_2$ are defined as
\begin{eqnarray}
I_1 ({\cal F}) &=&\int_0^{u_0} d\alpha_1 \big
[\frac{1}{u_0-\alpha_1}{\cal
F}(\alpha_1,1-u_0,u_0-\alpha_1)-\int_{u_0-\alpha_1}^{1-\alpha_1}
d\alpha_3 \frac{{\cal F}(\alpha_1,1-\alpha_1-\alpha_3,\alpha_3)}
{\alpha_3^2}\big ]\; , \\
I_2({\cal F}) &=& -\int_0^{1-u_0} d\alpha_3 \frac{{\cal F}(u_0,
1-u_0-\alpha_3, \alpha_3)}{\alpha_3} +\int_0^{u_0} d\alpha_1 \frac
{F(\alpha_1,1-u_0,u_0-\alpha_1)}{u_0-\alpha_1} \; ,
\end{eqnarray}
where ${\cal F}$ is one of the twist-four light cone amplitudes
$\varphi_{\|},~ \varphi_{\bot},~ {\tilde \varphi}_{\|}, ~{\tilde
\varphi}_{\bot}$.

In Eq.(\ref{sr1}), $M^2=\frac{M_1^2M_2^2}{M_1^2+M_2^2}$,
$u_0=\frac{M_1^2}{M_1^2+M_2^2}$. Since there exists an overlapping
working window for the two Borel parameters $M_1^2, M_2^2$, it's
convenient to let $M_1^2=M_2^2$, i.e., $u_0=\frac 12$. The eta
meson light-cone wave functions are known quite well at $u_0=\frac
12$. Such a choice allows the clean subtraction of the continuum
contribution. We can simply introduce a threshold parameter $s_0$
and replace $M^2 e^{-\frac{m_c^2}{M^2}}$ with $M^2\left(
e^{-\frac{m_c^2}{M^2}}-e^{-\frac{s_0}{M^2}}\right)$ to subtract
the contribution from the continuum and excited states
\cite{braun}.

\begin{figure}[hbt]
\begin{center}
\scalebox{0.8}{\includegraphics{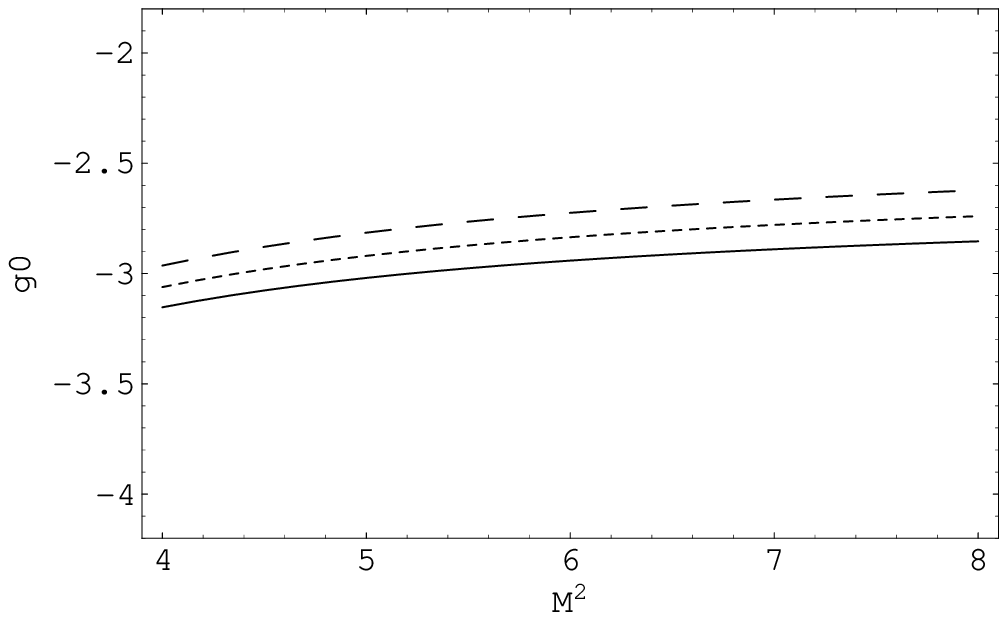}}
\end{center}
\caption{The variation of the coupling constant
$g_{D_{s0}D_s\eta}$ with $M^2$ (in unit of $\mbox{GeV}^2$). The
long-dashed, short-dashed and solid curves correspond to
$s_0=6.0,~ 6.25,~ 6.5~\mbox{GeV}^2$ respectively.}\label{fig1}
\end{figure}

In the numerical analysis, the values of parameters in the above
sum rule are: $m_{D_{s0}}=2.317$ GeV, $m_{D_s}=1.968$ GeV, $m_c$
(1 GeV)=1.35 GeV, $f_{D_s}=(266\pm 32)$ MeV \cite{pdg},
$f_{D_{s0}}=(225\pm 25)$ MeV \cite{Colangelo3}. For $f_{D_s}$ and
$f_{D_{s0}}$ we use the central values. Values of the other
parameters are given in the Appendix. The variation of
$g_{D_{s0}D_s\eta}$ with $M^2$ for the different $s_0$ is shown in
Fig. \ref{fig1}.

In the working region of the Borel parameter $5~\mbox{GeV}^2 < M^2
<7 ~\mbox{GeV}^2$, we get
\begin{equation}
-3.02<g_{D_{s0}D_s\eta}<-2.66 \; ,
 \end{equation}
where the uncertainty arises from the variation of $M^2$ and
$s_0$. Numerically, the twist-three term $\varphi_p$ has the
largest contribution to the sum rule.

The pionic decay of $D_{sJ}(2317)$ occurs through $\eta-\pi^0$
mixing, which is described by the isospin violating piece in the
chiral lagrangian
\begin{equation}
{\cal{L}}_m=\frac{m_{\pi}^2 f^2}{4(m_u+m_d)}\mbox{Tr}(\xi m_q
\xi+\xi^{\dagger} m_q \xi^{\dagger})~,
\end{equation}
where $\xi= \mbox{exp}(i\tilde{\pi}/f_\pi)$,  $\tilde{\pi}$ the
light meson octet and $m_q$ is the light quark mass matrix. The
mass difference between up and down quarks induces the
$\eta-\pi^0$ mixing with a suppression factor around
$\frac{m_d-m_u}{m_s-\frac{m_u+m_d}{2}}$. Finally the strong decay
width reads
\begin{equation}
\Gamma(D_{sJ}(2317)\rightarrow
D_s\pi^0)=\frac{3}{144\pi}g_{D_{s0}D_s\eta}^2
(\frac{m_d-m_u}{m_s-\frac{m_u+m_d}{2}})^2|\vec{p_1}|\; .
\end{equation}
Numerically we have
\begin{equation}
\Gamma(D_{sJ}(2317)\rightarrow D_s\pi^0)=(34-44)~\mbox{keV}\;.
\end{equation}

\section{$D_{sJ}(2460)\rightarrow D^{*}_s+\eta\rightarrow
D^{*}_s+\pi^0$}\label{sec3}

For $D_{sJ}(2460)$ decay, we define the following matrix element
\begin{equation}
\langle \eta(q)D_s^{*}(p)| D_{s1}^{'}(p+q)\rangle =
m_{D_{s1}^{'}}g_{D_{s1}^{'}D_s^{*}\eta}\eta^{\mu}\epsilon^{*}_{\mu}\;,
\end{equation}
where $\eta_{\mu}$ and $\epsilon_{\mu}$ are the polarization
tensors for the $1^+$ and $1^-$ states $D_{s1}^{'}, D_s^{*}$
respectively. We start from the correlation function
\begin{equation}
F_{\mu\nu}(p^2,(p+q)^2)=i \int d^4x \; e^{i p \cdot x} \langle
\eta(q) | T[J^\dagger_{\mu}(x) J^A_{\nu}(0)] |0\rangle
\end{equation}
where $J_{\mu}(x)=\bar c(x)\gamma_{\mu} s(x)$ and
$J^A_{\nu}(x)=\bar c(x) \gamma_{\nu}\gamma_5 s(x)$. At the hadron
level, we have
\begin{equation}
F_{\mu\nu}(p^2,(p+q)^2)=\frac{m_{D_{s1}^{'}}^2m_{D_s^{*}}f_{D_{s1}^{'}}f_{D_s^{*}}g_{D_{s1}^{'}D_s^{*}\eta}}
{(m_{D_s}^2-p^2)(m_{D_{s0}}^2-(p+q)^2)}(g_{\mu\nu}+
\frac{m_{D_{s1}^{'}}^2-m_{D_s^{*}}^2}{2m_{D_{s1}^{'}}^2m_{D_s^{*}}^2}q_{\mu}p_{\nu}+\ldots)\;,
\end{equation}
where we have kept the $g_{\mu\nu}$ and $q_{\mu}p_{\nu}$
structures. The decay constants $f_{D_{s1}^{'}}$ and $f_{D_s^{*}}$
are defined as
\begin{eqnarray}
\langle 0 | J_{\mu}^{+} | D_s^{*}\rangle&=& f_{D_s^{*}} m_{D_s^{*}}\epsilon_{\mu}\;,\\
\langle D_{s1}^{'} | J_{\nu}^A | 0\rangle&=&
f_{D_{s1}^{'}}m_{D_{s1}^{'}}\eta^{*}_{\nu}\; .
\end{eqnarray} \noindent

Following the same procedure as in Section \ref{sec2}, we obtain a
sum rule from the $g_{\mu\nu}$ structure
\begin{eqnarray}\label{q3}\nonumber
f_{D_{s1}^{'}}f_{D_s^{*}}g_{D_{s1}^{'}D_s^{*}\eta}&=&\frac{1}{m_{D_{s1}^{'}}^2m_{D_s}^{*}}
e^{\frac{m_{D_{s1}^{'}}^2+m_{D_s^{*}}^2}{2M^2}} \Bigg\{M^2
[e^{-\frac{m_c^2}{M^2}} - e^{-\frac{s_0}{M^2}}]
\Big[-\frac{1}{2}M^2F_{\eta}\varphi^{'}_{\eta}(u_0)
-m_cF_{\eta}\mu_{\eta}\varphi_p({u_0})\\\nonumber
 &&+\frac 18F_{\eta}m^2_{c}A^{'}(u_0)
+\frac 12 F_{\eta}m^2_{\eta}(2I_1(\varphi_{\parallel})-
I_2(\varphi_{\parallel}
+2\tilde{\varphi}_{\perp}-\tilde{\varphi}_{\|}))\big]\\
&& +e^{-\frac{m_c^2}{M^2}}\Big[\frac 18 F_{\eta} m_c^4A^{'}[u_0]
+F_{\eta}m_{\eta}^2m^2_{c}G(u_0)\Big]\Bigg\}_{u_0=1/2}\; .
\end{eqnarray}

Similarly we can get a second sum rule from the $q_{\mu}p_{\nu}$
structure
\begin{eqnarray}\label{q4}\nonumber
f_{D_{s1}^{'}}f_{D_s^{*}}g_{D_{s1}^{'}D_s^{*}\eta}&=&\frac{2m_{{D_s}^{*}}}{m_{D_{s1}^{'}}^2-m_{D_s^{*}}^2}
e^{\frac{m_{D_{s1}^{'}}^2+m_{D_s^{*}}^2}{2M^2}} \Bigg\{M^2
[e^{-\frac{m_c^2}{M^2}} - e^{-\frac{s_0}{M^2}}]
[-F_{\eta}\varphi_{\eta}(u_0)]
\\\nonumber
&& +e^{-\frac{m_c^2}{M^2}}\Big[\frac
13m_cF_{\eta}\mu_{\eta}\varphi_{\sigma}({u_0})+\frac 14 m_c^2
F_{\eta}(1+\frac{m_c^2}{M^2})A(u_0)-2F_{\eta}m_{\eta}^2u_0G(u_0)\\
&& +F_{\eta}m_{\eta}^2(I_3(\varphi_{\|}
+2\tilde{\varphi}_{\bot}-\tilde{\varphi}_{\|})
-2I_4(2\varphi_{\perp}+\varphi_{\|}))\Big]\Bigg\}_{u_0=1/2}\;.
\end{eqnarray}

The functions $I_3$ and $I_4$ in Eqs. (\ref{q3}) and (\ref{q4})
are defined as
\begin{eqnarray}
 I_3 ({\cal F}) &=&\int_0^{u_0} d\alpha_1
\int_{u_0-\alpha_1}^{1-\alpha_1} d\alpha_3 \frac {{\cal
F}(\alpha_1,
1-\alpha_1-\alpha_3, \alpha_3) }{\alpha_3}\; ,\\
I_4 ({\cal F}) &=&\int_0^{u_0} d\alpha_1
\int_{u_0-\alpha_1}^{1-\alpha_1} d\alpha_3 \frac {u_0-\alpha_1
}{\alpha^2_3}{\cal F}(\alpha_1, 1-\alpha_1-\alpha_3, \alpha_3)\; .
\end{eqnarray}

Unfortunately the sum rule Eq. (\ref{q4}) is very unstable. There
is no working window for the Borel parameter $M^2$. In the
following we focus on the sum rule Eq. (\ref{q3}). We use
$m_{D_{s1}^{'}}=2.460 ~\mbox{GeV},~ m_{D_s^{*}}=2.112 ~\mbox{GeV}$
\cite{pdg}, $f_{D_{s1}^{'}}\simeq f_{D_{s0}}$, $f_{D_s^{*}}\simeq
f_{D_s}$ \cite{colangelo2}. The variation of
$g_{D_{s1}^{'}D_s^{*}\eta}$ with $M^2$ is presented in Fig. 2.

\begin{figure}[hbt]
\begin{center}
\scalebox{0.8}{\includegraphics{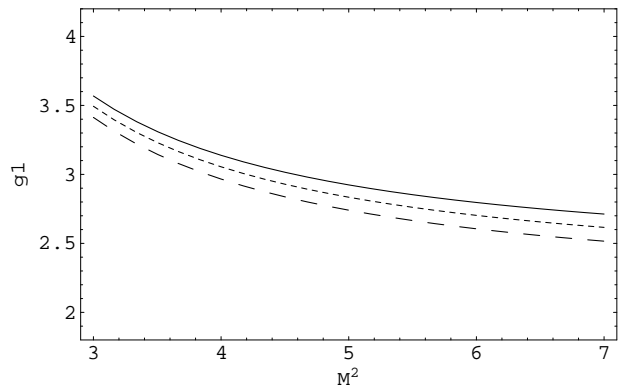}}
\end{center}
\caption{The variation of the coupling constant
$g_{D_{s1}^{'}D_s^{*}\eta}$ with $M^2$ (in unit of
$\mbox{GeV}^2$). The long-dashed, short-dashed and solid curves
correspond to $s_0=6.25, ~6.5, ~6.75~\mbox{GeV}^2$ respectively.
}\label{fig2}
\end{figure}

In the working window of Borel parameter $4 ~\mbox{GeV}^2 < M^2 <6
~\mbox{GeV}^2$, we have
\begin{equation}
2.61<g_{D_{s1}^{'}D_s^{*}\eta}<3.14 \; .
\end{equation}
The contribution from $\varphi_p$ term is also very important
numerically. The pinonic decay width reads
\begin{equation}
\Gamma(D_{sJ}(2460)\rightarrow
D^{*}_s+\pi^0)=\frac{g_{D_{s1}^{'}D_s^{*}\eta}^2}{144\pi}
(2+\frac{(m_{D_{s1}^{'}}^2+m_{D_s^{*}}^2)^2}{4m_{D_{s1}^{'}}^2m_{D_s^{*}}^2})
(\frac{m_d-m_u}{m_s-\frac{m_u+m_d}{2}})^2|\vec{p_1}|\; .
\end{equation}
Finally we have
\begin{equation}
\Gamma(D_{sJ}(2460)\rightarrow D^{*}_s+\pi^0)=(35-51) ~\mbox{keV}
\;.
\end{equation}

\section{Discussion}\label{sec4}

The strong decay widths of $D_{sJ}(2317)\rightarrow D_s+\pi^0$ and
$D_{sJ}(2460)\rightarrow D^{*}_s+\pi^0$ have been calculated by
several groups. Their results are collected in Table \ref{table1}
together with ours. The first five calculations assume $c\bar s$
picture while the last two use composite non-$c\bar s$ pictures.
The decay width of $D_{sJ}(2317)$ is roughly the same as that of
$D_{sJ}(2460)$ from all approaches. The $1/m_c$ correction is
expected to modify the small values in the second column from
vector dominance model in the heavy quark limit \cite{colangelo2}.

\begin{table}[h]
\caption{\baselineskip 15pt Strong decay widths (in keV) of
$D_{sJ}(2317)$ and $D_{sJ}(2460)$ from various theoretical
approaches.} \label{table1}
    \begin{center}
    \begin{tabular}{c c c c c c c c c}
\hline  & LCQSR &\cite{colangelo1} &
\cite{Bardeen} & \cite{godfrey} &\cite{fayy}&\cite{tetra1}&\cite{ishida}&\\
\hline
$D^*_{sJ}(2317)\rightarrow D_{s}\pi^0$&  34-44   & $7\pm 1  $& 21.5  &$\simeq 10 $& 16 & 10-100&$150\pm 70$\\
$D_{sJ}(2460)\rightarrow D_{s}^{*}\pi^0$& 35-51  & $7\pm 1  $& 21.5  &$\simeq 10 $& 32 &       &$150\pm 70$\\
\hline
\end{tabular}
\end{center}
\end{table}

\begin{table}[h]
\caption{\baselineskip 15pt Comparison between experimental ratio
of $D_{sJ}(2317, 2460)$ radiative and strong decay widths and
theoretical predictions from LCQSR based on Ref. \cite{colangelo2}
and this work.} \label{table2}
    \begin{center}
    \begin{tabular}{cccc|c}
\hline & Belle & Babar   & CLEO \cite{cleo} & LCQSR\\ \hline
$\frac{\Gamma \left( D^*_{sJ}(2317) \rightarrow D_{s}^{\ast
}\gamma \right) }{ \Gamma \left( D^*_{sJ}(2317)\rightarrow
D_{s}\pi ^{0}\right) }$ & $<0.18$ \cite{belle2}&
 \ \hfill --- \hfill\   & $<0.059$& 0.13 \\
\hline $\frac{\Gamma \left(D_{sJ}(2460) \rightarrow D_{s}\gamma
\right) }{ \Gamma \left( D_{sJ}(2460)\rightarrow D_{s}^{\ast }\pi
^{0}\right) }$ &
\begin{tabular}{c}
$0.55\pm0.13\pm0.08$ \cite{belle2}\\
\end{tabular}&
\begin{tabular}{c}
$0.375\pm0.054\pm0.057$ \cite{babar5} \\
\end{tabular}
&$<0.49$ & 0.56\\
\hline $\frac{\Gamma \left( D_{sJ}(2460)\rightarrow
D_{s}^{\ast }\gamma \right) }{\Gamma \left( D_{sJ}(2460)\rightarrow D_{s}^{\ast }\pi ^{0}\right) }$
& $<0.31$ \cite{belle2}&\ \hfill --- \hfill\ &$<0.16$ & 0.02 \\
\hline  $\frac{\Gamma \left( D_{sJ}(2460)\rightarrow
D^*_{sJ}(2317)\gamma \right) }{\Gamma \left(
D_{sJ}(2460)\rightarrow
D_{s}^*\pi^0 \right) }$  & \ \hfill --- \hfill\ & $ < 0.23$   \cite{babar4}& $ < 0.58$ & 0.015 \\
\hline
\end{tabular}
\end{center}
\end{table}

The radiative decay widths of $D_{sJ}(2317, 2460)$ were calculated
using LCQSR in \cite{colangelo2}: $\Gamma (D_{sJ}(2317)\rightarrow
D^{*}_s+\gamma)=(4-6)~\mbox{keV},~
 \Gamma(D_{sJ}(2460)\rightarrow D_s \gamma)=(19-29)~\mbox{keV} ,~
 \Gamma(D_{sJ}(2460)\rightarrow D^{*}_s+\gamma)=(0.6-1.1) ~\mbox{keV},~
 \Gamma(D_{sJ}(2460)\rightarrow D_{sJ}(2317)+\gamma)=(0.5-0.8)~
 \mbox{keV}$. Experimentally only $D_{sJ}(2460)\rightarrow D_s \gamma$ has
been observed by Belle \cite{belle1,belle2} and Babar
\cite{babar2,babar5}. We have collected the experimental ratio of
radiative and strong decays of $D_{sJ}$ mesons together with the
central values of theoretical predictions from LCQSR based on Ref.
\cite{colangelo2} and present work in Table \ref{table2}. For
$D_{sJ}(2460)\rightarrow D_s \gamma)$, we get a range 0.37-0.83
for the ratio, consistent with both Belle and Babar's measurement.

In short summary, we have calculated the coupling constants
$g_{D_{s0}D_s\eta}$ and $g_{D_{s1}^{'}D_s^{*}\eta}$ in the
framework of LCQSR. Through the $\eta-\pi^0$ mixing we obtain
$\Gamma(D_{sJ}(2317)\rightarrow D_s\pi^0)=(34-44)~\mbox{keV}$ and
$\Gamma(D_{sJ}(2460)\rightarrow D_s^{*}\pi^0)=(35-51)~\mbox{keV}$.
These two widths are similar in magnitude, as expected from heavy
quark symmetry. The ratio between the radiative widths and strong
decay widths obtained in the same LCQSR framework is consistent
with Belle and Babar's most recent data, which strongly indicates
$D_{sJ}(2317)$ and $D_{sJ}(2460)$ are conventional $c\bar s$
mesons. In the future, B decays into $D_{sJ}$ mesons may also play
an important role in exploring these charming states
\cite{suzuki,chen,cheng,datta,huang,cheng2,barnes}.

\section{Acknowledgments}

This project was supported by the National Natural Science
Foundation of China under Grants 10375003 and 10421003, Ministry
of Education of China, FANEDD, Key Grant Project of Chinese
Ministry of Education (NO 305001) and SRF for ROCS, SEM. W.W.
thanks Jie Lu and Feng-Lin Wang for helpful discussions.

\section*{Appendix }
We use ${\bar q \Gamma q}$ to denote $({\bar u \Gamma u}+ {\bar d
\Gamma d} -2 {\bar s \Gamma s})/ \sqrt{6}$. Up to twist four, the
two- and three-particle light-cone wave functions of eta meson can
be written as \cite{ballA,ballB}:
\begin{eqnarray}\label{phieta}\nonumber
<\eta| {\bar q} (x) \gamma_{\mu} \gamma_5 q(0) |0>&=&-i f_{\eta}
q_{\mu} \int_0^1 du \; e^{iuqx} [\varphi_{\eta}(u) +{1\over 16}
m^2_\eta x^2 A(u)  ] \nonumber\\
&& -{i\over 2}f_\eta m_\eta^2
{q_\mu \over q x} \int_0^1 du \; e^{-iuqx}  B(u) \; ,\nonumber\\
<\eta| {\bar q} (x) i \gamma_5 q(0) |0>&=& f_\eta \mu_\eta
\int_0^1 du \; e^{iuqx} \varphi_P (u)  \; ,\nonumber\\
<\eta| {\bar q}(x) \sigma_{\mu \nu} \gamma_5 q(0) |0>&=& {i\over
6} f_\eta \mu_\eta (q_\mu x_\nu-q_\nu x_\mu) \int_0^1 du \;
e^{-iuqx}
\varphi_\sigma (u)  \; ,\nonumber\\
<\eta | {\bar q} (x) \sigma_{\alpha \beta} \gamma_5 g_s G_{\mu
\nu}(ux) q(x) |0>&=& i f_\eta \mu_\eta \eta_3 [(q_\mu q_\alpha
g_{\nu \beta}-q_\nu q_\alpha g_{\mu \beta}) -(q_\mu q_\beta g_{\nu
\alpha}-q_\nu q_\beta g_{\mu \alpha})]\nonumber \\
&&\int {\cal D}\alpha_i \; \varphi_{3 \eta} (\alpha_i)
e^{-iqx(\alpha_1+v \alpha_3)}  \; ,\nonumber \\
<\eta| {\bar q} (x) \gamma_{\mu} \gamma_5 g_s G_{\alpha \beta}(vx)
q(0) |0 >&=& f_{\eta}m_\eta^2 \Big[ q_{\beta} \Big( g_{\alpha
\mu}-{x_{\alpha}q_{\mu} \over q \cdot x} \Big) -q_{\alpha} \Big(
g_{\beta \mu}-{x_{\beta}q_{\mu} \over q \cdot x} \Big) \Big] \int
{\cal{D}} \alpha_i \varphi_{\bot}(\alpha_i)
e^{-iqx(\alpha_1 +v \alpha_3)}\nonumber \\
&&+f_{\eta} m_\eta^2 {q_{\mu} \over q \cdot x } (q_{\alpha}
x_{\beta}-q_{\beta} x_{\alpha}) \int {\cal{D}} \alpha_i
\varphi_{\|} (\alpha_i) e^{-iqx(\alpha_1 +v \alpha_3)}  \; ,\nonumber\\
<\eta|{\bar q} (x) \gamma_{\mu}  g_s \tilde G_{\alpha
\beta}(vx)q(x) |0>&=& i f_{\eta} m_\eta^2 \Big[ q_{\beta} \Big(
g_{\alpha \mu}-{x_{\alpha}q_{\mu} \over q \cdot x} \Big)
-q_{\alpha} \Big( g_{\beta \mu}-{x_{\beta}q_{\mu} \over q \cdot x}
\Big) \Big] \int {\cal{D}} \alpha_i \tilde
\varphi_{\bot}(\alpha_i)
e^{-iqx(\alpha_1 +v \alpha_3)}\nonumber \\
&&i f_{\eta} m_\eta^2  {q_{\mu} \over q \cdot x } (q_{\alpha}
x_{\beta}-q_{\beta} x_{\alpha}) \int {\cal{D}} \alpha_i \tilde
\varphi_{\|} (\alpha_i) e^{iqx(\alpha_1 +v \alpha_3)} \;,\nonumber
\end{eqnarray}
where the operator $\tilde G_{\alpha \beta}$  is the dual of
$G_{\alpha \beta}$: $\tilde G_{\alpha \beta}= {1\over 2}
\epsilon_{\alpha \beta \delta \rho} G^{\delta \rho} $; ${\cal{D}}
\alpha_i$ is defined as ${\cal{D}} \alpha_i =d \alpha_1 d \alpha_2
d \alpha_3 \delta(1-\alpha_1 -\alpha_2 -\alpha_3)$ and
$f_{\eta}\simeq 1.2f_{\pi}=0.156~\mbox{GeV},~ \eta_3=0.013
,~m_s(1~\mbox{GeV})=0.125~\mbox{GeV} ,~ m_{\eta}=0.548
~\mbox{GeV},~ \mu_{\eta}=\frac{m_{\eta}^2}{m_s}=2.4 ~\mbox{GeV}$
\cite{pdg}.

The distribution amplitudes $\varphi_{\eta}$ etc can be
parameterized as
\begin{eqnarray}
\varphi_{\eta}(u) &=&
6u\bar{u}(1+a_2C^{3/2}_2(\zeta)+a_4C^{3/2}_4(\zeta))\; ,\nonumber\\
\phi_p(u) &=& 1+(30\eta_3-\frac 52 \rho_{\eta}^2)
C^{1/2}_2(\zeta)+(-3\eta_3\omega_3-\frac{27}{20}\rho_{\eta}^2-\frac
{81}{10}\rho_{\eta}^2a_2)
C^{1/2}_4(\zeta)  \; ,\nonumber\\
\phi_{\sigma}(u)&=&
6u(1-u)\big\{1+(5\eta_3-\frac{1}{2}\eta_3\omega_3-\frac{27}{20}\rho_{\eta}^2-\frac
{3}{5}\rho_{\eta}^2a_2)\big\}
C^{3/2}_2(\zeta)  \; ,\nonumber\\
g_{\eta}(u)&=& 1+\{1+\frac{18}{7}a_2+60\eta_3+\frac{20}{3}\eta_4\}
C^{1/2}_2(\zeta)
+\{-\frac{9}{28}a_2-6\eta_3\omega_3\}C^{1/2}_4(\zeta) \; ,\nonumber\\
{\mathbb A}(u) & = & 6u\bar u \{ \frac{16}{15} +
\frac{24}{35}a_2 + 20 \eta_3 + \frac{20}{9}\eta_4 \nonumber\\
&& + (-\frac{1}{15} + \frac{1}{16}-\frac{7}{27} \eta_3 \omega_3 -
\frac{10}{27}\eta_4 ) C_2^{3/2}(\xi) + ( -\frac{11}{210} a_2 -
\frac{4}{135}\eta_3\omega_3 )C_4^{3/2}(\xi) \} \nonumber\\
&& + (-\frac{18}{5} a_2 + 21\eta_4\omega_4 ) \{ 2
  u^3 (10-15 u + 6 u^2)\ln u + 2\bar u^3 (10-15\bar u + 6 \bar u^2)
  \ln\bar u \nonumber \\
& & + u \bar u (2 + 13u\bar u)\} \; ,\nonumber\\
\varphi_{3\eta}(\alpha_1,\alpha_2,\alpha_3)&=& 360 \eta_3
\alpha_1\alpha_2\alpha_3^2 \{1+\frac{1}{2}\omega_3(7\alpha_3-3)\}
 \; ,\nonumber\\
\varphi_{\parallel}(\alpha_1,\alpha_2,\alpha_3)&=& 120
\alpha_2\alpha_1
\alpha_3(a_{10} (\alpha_1-\alpha_2) \; ,\nonumber\\
\varphi_{\perp}(\alpha_1,\alpha_2,\alpha_3)&=&
  30\alpha_3^2(\alpha_2-
\alpha_1)[h_{00}+ h_{01}\alpha_3
+\frac{1}{2}h_{10}(5\alpha_3-3)  \; ,\nonumber\\
\tilde{\varphi}_{\parallel}(\alpha_1,\alpha_2,\alpha_3)&=& 120
\alpha_1\alpha_2 \alpha_3(v_{00}+
v_{10}(3\alpha_3-1) \; , \nonumber\\
\tilde{\varphi}_{\perp}(\alpha_1,\alpha_2,\alpha_3)&=&
-30\alpha_3^2\{h_{00} \overline{\alpha}_3+
h_{01}[\alpha_g\overline{\alpha}_3-6\alpha_1\alpha_2]
+h_{10}[\alpha_3\overline{\alpha}_3-\frac{3}{2}
(\alpha_1^2+\alpha_2^2)] \; ,\nonumber
\end{eqnarray}
where $\bar{u} \equiv 1-u \; , \zeta \equiv 2u-1 \; ,
\overline{\alpha}=1-\alpha$. $C^{3/2,1/2}_{2,4}(\zeta)$ are
Gegenbauer polynomials. Here $g_{\eta}(u)=B(u)+\varphi_{\eta}(u)$.
$\rho_{\eta}^2$ gives the mass correction and are defined as
$\rho_{\eta}^2=\frac {m_s^2}{m_{\eta}^2}$.  $a_{ij}$, $v_{ij}$ and
$h_{ij}$ are related to hadronic matrix elements $\eta_4$,
$\omega_4$ and $a_2$ as
$$
\begin{array}{r@{~=~}l@{\,,\quad}r@{~=~}l@{\,,\quad}r@{~=~}l}
a_{10} & \frac{21}{8}\eta_4\omega_4-\frac{9}{20}a_2 & v_{10} &
\frac{21}{8}\eta_4\omega_4 &
v_{00} & -\frac{1}{3}\eta_4\,, \\[5pt]
h_{01} & \frac{7}{4}\eta_4\omega_4-\frac{3}{20}a_2 & h_{10} &
\frac{7}{2}\eta_4\omega_4+\frac{3}{20}a_2 & v_{00} &
-\frac{1}{3}\eta_4 \,.
\end{array}
$$
The values of $a_2$ et al are: $a_2=0.115,~ a_4=-0.015,~
\eta_3=0.013,~ \omega_3=-3,~ \eta_4=0.5,~ \omega_4=0.2$. All of
them are scaled at $\mu=1~\mbox{GeV}$.

\end{document}